\documentclass[runningheads]{llncs}
\usepackage[T1]{fontenc}
\usepackage{graphicx}
\usepackage{amsmath}    
\usepackage{amssymb}    
\usepackage{mathtools}  
\usepackage{booktabs}   

\usepackage{tikz}
\usetikzlibrary{arrows.meta, positioning, decorations.pathreplacing}
\usepackage{tikz}
\usetikzlibrary{positioning}
\begin{document}
\pagestyle{empty}

\title{Quantum Approaches to the Minimum Edge Multiway Cut Problem}

\author{
Ali Abbassi\inst{1,2} \and
Yann Dujardin\inst{1} \and
Eric Gourdin\inst{1} \and
Philippe Lacomme\inst{3} \and
Caroline Prodhon\inst{2}
}

\authorrunning{A. Abbassi et al.}

\institute{
Orange Research, Chatillon, France
\and
LIST3N, University of Technology of Troyes, Troyes, France
\and
LIMOS - UMR CNRS 6158, ISIMA - Université Clermont Auvergne, France
\\
\email{ali.abbassi@{utt.fr,orange.com}, yann.dujardin@orange.com, eric.gourdin@orange.com, philippe.lacomme@isima.fr, caroline.prodhon@utt.fr}
}
\maketitle              
\begin{abstract}
We investigate the minimum edge multiway cut problem, a fundamental task in evaluating the resilience of telecommunication networks. This study benchmarks the problem across three quantum computing paradigms: quantum annealing on a D-Wave quantum processing unit, photonic variational quantum circuits simulated on Quandela’s Perceval platform, and IBM’s gate-based Quantum Approximate Optimization Algorithm (QAOA). We assess the comparative feasibility of these approaches for early-stage quantum optimization, highlighting trade-offs in circuit constraints, encoding overhead, and scalability. Our findings suggest that quantum annealing currently offers the most scalable performance for this class of problems, while photonic and gate-based approaches remain limited by hardware and simulation depth. These results provide actionable insights for designing quantum workflows targeting combinatorial optimization in telecom security and resilience analysis.

\end{abstract}

%
%
%
\section{Introduction}

The \emph{Minimum Edge Multiway Cut} (MEMC) problem arises in tasks where specific terminal nodes must be separated to satisfy specific resilience criteria for a network  or unauthorized communication~\cite{Costa2005MulticutSurvey}. Given an undirected graph and a set of $k$ terminals, the objective is to identify a minimum-weight set of edges whose removal ensures that no two terminals lie in the same connected component. For $k=2$, the problem reduces to a classical min-cut, solvable in polynomial time via max-flow algorithms~\cite{Garg1996Approx}. For $k \geq 3$, the problem is NP-hard~\cite{Dahlhaus1994Multiterminal}.
We evaluate the MEMC problem using three quantum computing paradigms: (i) quantum annealing on D-Wave hardware, (ii) the Quantum Approximate Optimization Algorithm (QAOA) implemented with IBM Qiskit, and (iii) variational photonic quantum circuits simulated using Quandela’s Perceval platform. Each paradigm operates on a shared QUBO formulation of MEMC, enabling consistent problem modeling across solvers~\cite{CruzSantos2023DWaveQUBO}.

The QUBO model is mapped to an Ising formulation for execution on the D-Wave platform and is employed as the cost Hamiltonian for QAOA and photonic variational circuits. D-Wave serves as a natural baseline due to its native support for QUBO-based quantum annealing and available hybrid solvers~\cite{DWaveDocs2024}. We report numerical results on small graph instances, including annealing behavior, QAOA convergence trends, and bitstring sampling accuracy in photonic circuits. 

This  study aims to characterize the behavior of different quantum optimization approaches on a unified problem encoding, highlighting trade-offs in scalability, model expressiveness, and hardware  compatibility~\cite{Alex}. The paper is structured as follows: Section~2 defines the MEMC problem and recalls key complexity results. Section~3 presents the QUBO formulation used across all quantum paradigms. Section~4 introduces the three quantum optimization approaches: quantum annealing, QAOA, and photonic variational circuits. Section~5 reports numerical results obtained on representative instances. Section~6 concludes with a discussion of current limitations and directions for future work.

\section{Problem Statement}

We consider the \emph{Minimum Edge Multiway Cut} (MEMC) problem, a classical NP-hard graph partitioning problem with applications in network interdiction, secure communications, and containment strategies~\cite{Magnouche2020MTVS}.

\medskip
\noindent\textbf{Input.} An undirected connected graph \( G = (V, E) \), a non-negative cost function \( C : E \to \mathbb{R}_{\geq 0} \), and a set of \( k \geq 2 \) terminals \( T = \{t_1, \ldots, t_k\} \subseteq V \).

\medskip
\noindent\textbf{Goal.} Find a minimum-cost set of edges \( F \subseteq E \) such that in the graph \( G' = (V, E \setminus F) \), no two terminals lie in the same connected component.
\medskip

The problem generalizes the classic \emph{minimum $s$–$t$ cut} (when \( k = 2 \)), solvable in polynomial time via max-flow algorithms~\cite{Garg1996Approx}. For \( k \geq 3 \), the problem becomes NP-hard~\cite{Dahlhaus1994Multiterminal}, and remains hard even under restrictions such as bounded degree or treewidth~\cite{Calinescu2003Multicuts}\cite{PapadopoulosPermutation}. While fixed-parameter tractable algorithms exist for certain parameters~\cite{Bousquet2011Multicut}\cite{Marx2011MulticutFPT}, the decision variant—testing whether a cut of cost at most \( K \) exists—is NP-complete in general. Greedy heuristics offer a \( 2 - \frac{2}{k} \) approximation ratio~\cite{Dahlhaus1994Multiterminal}, but exact solutions for larger instances often require exponential-time methods or relaxation techniques.
We consider the optimization variant and reformulate it as a Quadratic Unconstrained Binary Optimization (QUBO), which serves as a common ground for deploying quantum solvers.

\section{QUBO formulation}
We adopt the QUBO formulation for the MEMC problem from~\cite{Heidari2022QUBO}, designed for efficient embedding on quantum hardware. Given a graph $G=(V,E)$, each vertex $u \in V$ is assigned to one of the terminals $t \in T$ via binary variables $x_{u,t} \in \{0,1\}$, where $x_{u,t} = 1$ if node $u$ is assigned to terminal $t$. This defines a partition of the graph vertices. We hence consider the following Hamiltonian:
\begin{equation}
\begin{aligned}
    H(x) =\ & \alpha \left( \sum_{u \in V} \left( 1 - \sum_{t \in T} x_{u,t} \right)^2 
    + \sum_{\substack{t, t' \in T \\ t \neq t'}} x_{t,t'} \right) \\
    & + \sum_{\{u,v\} \in E} \sum_{t \in T} \sum_{\substack{t' \in T \\ t \neq t'}} 
    C(\{u, v\}) x_{u,t} x_{v,t'}
\end{aligned}
\label{eq:qubo_resized}
\end{equation}
The first term ensures that each node is assigned to exactly one terminal, and the second term prevents conflicting terminal overlaps
(no terminal is assigned to another terminal). 

The final term accumulates the cost of cutting edges that connect vertices assigned to different terminals. The scalar $\alpha$ regulates the strength of the constraint terms relative to the objective.

\section{Quantum Optimization}

We describe the different quantum optimization workflows considered in this study and briefly recall their theoretical soundness and encoding strategies~\cite{Abbass}\cite{Alex}.

\subsection{Quantum Annealing}

Quantum annealing (QA) is a continuous-time optimization method based on the quantum adiabatic theorem~\cite{Farhi2000QAA}. The system evolves under a time-dependent Hamiltonian
\begin{equation}
    H(t) = A(t) H_0 + B(t) H_p, \quad t \in [0,T],
\end{equation}
where $H_0$ is a transverse-field driver and $H_p$ encodes the cost function. Provided $A(t), B(t)$ vary slowly enough and the spectral gap remains non-negligible, the system ends near the ground state of $H_p$.

Our QUBO formulation is mapped to an Ising energy function via $x_i = (1 - z_i)/2$ with $z_i \in \{-1,1\}$~\cite{Glover2019QUBOTutorial}:
\begin{equation}
    E(z) = \sum_{i<j} J_{ij} z_i z_j + \sum_i h_i z_i,
\end{equation}
where $J_{ij} = \frac{1}{4} Q_{ij}$ and $h_i = -\frac{1}{4} \sum_j (Q_{ij} + Q_{ji})$. This defines an Ising spin glass whose ground state solves the original problem.

The Ising cost function is promoted to a diagonal quantum Hamiltonian:
\begin{equation}
    H_p = \sum_{i<j} J_{ij} Z_i Z_j + \sum_i h_i Z_i,
\end{equation}
where $Z_i$ is the Pauli-$Z$ operator on qubit $i$. On D-Wave systems, this Hamiltonian is implemented using programmable couplings and local fields over a sparse hardware graph~\cite{DWaveDocs2024}. Minor embedding is used to satisfy hardware connectivity constraints by linking physical qubits into chains, enabling annealing toward a minimum-energy configuration that corresponds to low cost solution.

\subsection{Quantum Approximate Optimization Algorithm}

The Quantum Approximate Optimization Algorithm (QAOA)~\cite{farhi2014qaoa} is a variational quantum algorithm for solving discrete optimization problems using alternating Hamiltonian dynamics on shallow circuits. It operates on a cost Hamiltonian $H_C$, derived from the QUBO model, and a mixer Hamiltonian $H_M$. The QAOA$_p$ ansatz prepares the state
\[
|\psi(\vec{\gamma}, \vec{\beta})\rangle = \prod_{\ell=1}^{p} e^{-i \beta_\ell H_M} e^{-i \gamma_\ell H_C} |\psi_0\rangle,
\]
where $\vec{\gamma}, \vec{\beta} \in \mathbb{R}^p$ are tunable parameters and $|\psi_0\rangle = |+\rangle^{\otimes n}$ is the initial uniform superposition. In practice, $H_C$ is implemented using single-qubit $R_Z$ gates; the standard mixer $H_M = \sum_i X_i$ corresponds to global $R_X$ rotations.

The parameters are optimized by minimizing the energy expectation value $\langle \psi(\vec{\gamma}, \vec{\beta}) | H_C | \psi(\vec{\gamma}, \vec{\beta}) \rangle$ using classical optimizers such as COBYLA. The QAOA structure can be interpreted as a Trotterized approximation of evolution under $H = H_C + H_M$, with a “bang-bang” control profile~\cite{brady2021bang}.

While standard QAOA explores the full Hilbert space, constrained extensions like the Quantum Alternating Operator Ansatz (QAOA-A)~\cite{hadfield2019quantum} use custom mixers to enforce problem constraints. These were not used in our experiments; instead, constraints are embedded directly in the QUBO energy landscape.


\subsection{Variational Circuits on Photonic Quantum Computers}

Photonic variational quantum circuits offer an alternative paradigm for solving combinatorial optimization problems using linear optical components~\cite{Heurtel2022Perceval}~\cite{Bombardelli2025Foundations}. These circuits operate in the discrete-variable regime, where Fock states—photon number states—form the computational basis. Unitary transformations are implemented via passive interferometers built from beam splitters and phase shifters. Classical optimization is used to adjust circuit parameters to minimize a QUBO-derived cost function.

Each Fock state $|\vec{n}\rangle = |n_0, n_1, \ldots, n_{M-1}\rangle$ encodes the number of photons in $M$ modes, with $\sum_j n_j = P$ photons in total. For binary optimization, encodings such as dual-rail and parity are used. In dual-rail encoding, a logical qubit spans two modes:
\[
|0\rangle = |1\rangle_0 \otimes |0\rangle_1, \quad
|1\rangle = |0\rangle_0 \otimes |1\rangle_1.
\]
We use a parity-based mapping \cite{Bradler2021Bosonic}, where each bit is computed as $x_i = n_i \bmod 2$, enabling direct evaluation of QUBO objectives: $H(x) = x^\top Q x$.

The circuit applies a unitary transformation $U(\boldsymbol{\theta}, \boldsymbol{\phi})$ to the input Fock state $|\vec{n}_{\mathrm{in}}\rangle$. This unitary is decomposed into parameterized beam splitters and phase shifters:
\[
\text{BS}(\theta, \phi) =
\begin{pmatrix}
\cos(\theta) & -e^{i\phi} \sin(\theta) \\
e^{-i\phi} \sin(\theta) & \cos(\theta)
\end{pmatrix}, \quad
\text{PS}(\varphi) =
\begin{pmatrix}
1 & 0 \\
0 & e^{i\varphi}
\end{pmatrix}.
\]
These gates evolve the quantum state to:
\[
|\psi_{\mathrm{out}}\rangle = U(\boldsymbol{\theta}, \boldsymbol{\phi}) |\vec{n}_{\mathrm{in}}\rangle.
\]

Measurements in the Fock basis yield output configurations $\vec{n}_{\mathrm{out}}$, which are mapped to bitstrings $x \in \{0,1\}^N$ via parity. The QUBO energy is estimated as:
\[
\mathbb{E}_{x \sim p(\boldsymbol{\theta}, \boldsymbol{\phi})}[H(x)] = \sum_x p(x) \cdot x^\top Q x.
\]
Minimizing this expected value over $(\boldsymbol{\theta}, \boldsymbol{\phi})$ forms the variational optimization loop. We simulate small MEMC instances using \texttt{Perceval}~\cite{Heurtel2022Perceval} under idealized conditions and evaluate the circuit's ability to sample near-optimal bitstrings.


\section{Numerical Analysis}

Preliminary results confirm that \emph{small-scale instances} of the MEMC problem can be solved to optimality across all three quantum workflows—quantum annealing (QA), gate-based QAOA, and photonic variational circuits (VQC). In cases with $k = 2$ and $|V| \leq 10$, all platforms return the optimal solution, demonstrating that constrained combinatorial problems are tractable in low-dimensional quantum settings.

\begin{table}[htbp]
\centering
\scriptsize
\renewcommand{\arraystretch}{1.05}
\caption{Numerical summary}
\label{tab:backend_summary}
\begin{tabular}{|p{1.2cm}|p{2.5cm}|p{3.2cm}|p{5.2cm}|}
\toprule
\textbf{Backend} &
\shortstack{\scriptsize Small instances \\ \scriptsize ($k=2$, $|V|\leq 10$)} &
\shortstack{\scriptsize Larger instances \\ \scriptsize ($k=2$, $10<|V|<70$)} &
\scriptsize Convergence / notes \\
\midrule
D-Wave QA &
Optimal (100\%) &
Gap $\approx 20$--$40$\% (topology-dependent) &
Chain breaks observed, avg.\ chain length $\sim 8$; hybrid solver recommended for stability. \\
\midrule
D-Wave Hybrid &
Optimal (100\%) &
Near-optimal; typical gap $\lesssim 10$--$15$\% &
Longer runtime; classical post-processing mitigates embedding effects. \\
\midrule
QAOA (sim.) &
Optimal bitstring sampled with $\sim 40$\% probability at $p=1$; no improvement for larger iterations. &
Gap $\sim 15$--$30$\% at fixed $p$; no systematic improvement for larger $p$ (barren plateau) &
Converges in $\sim$10--20 iterations; decreasing probability of sampling best bitstring at $p=1$, effect of BP for medium to large instances. \\
\midrule
Photonic (sim.) &
Optimal sampled with $>90$\% probability. &
Gap $\sim 20$--$40$\%; infeasible to simulate larger systems with SLOS backend &
$\sim$300--1500 parameter updates; Nelder--Mead more stable than COBYLA. \\
\bottomrule
\end{tabular}
\end{table}

\begin{figure}[hbtp]
\centering

\begin{minipage}[t]{0.48\textwidth}
    \centering
    \includegraphics[width=\linewidth]{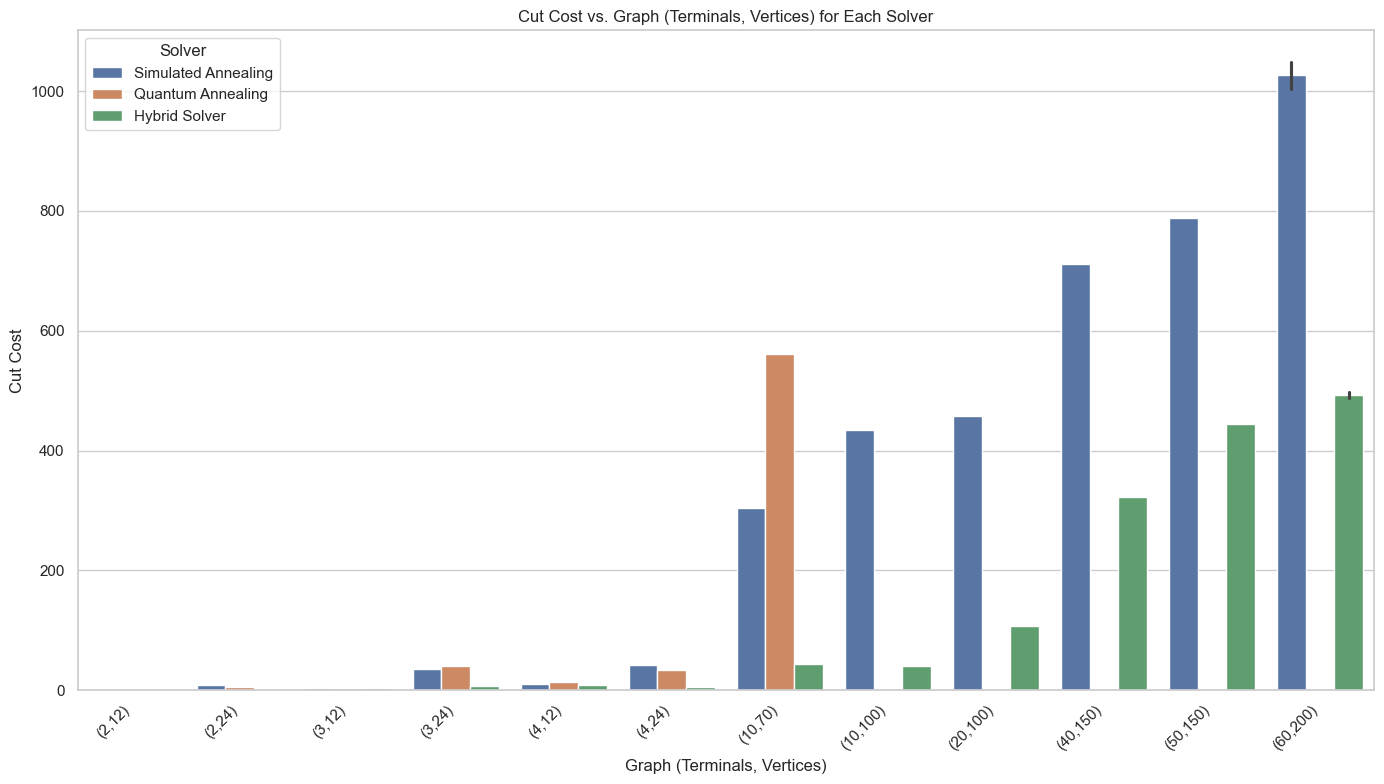}
    \caption{Hybrid vs Quantum Annealing Cut Cost}
    \label{fig:fig_qa_cost}
\end{minipage}
\hfill
\begin{minipage}[t]{0.48\textwidth}
    \centering
    \includegraphics[width=\linewidth]{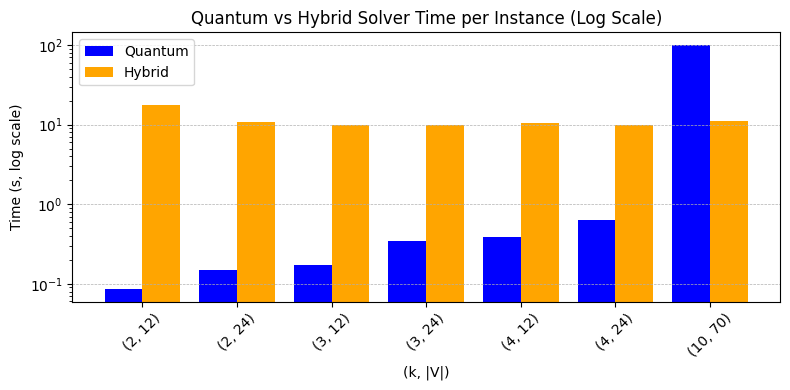}
    \caption{D-Wave Runtime vs Instance Size}
    \label{fig:fig_qa_runtime}
\end{minipage}
\end{figure}

Figure~\ref{fig:fig_qa_cost} compares cut costs obtained by simulated annealing, quantum annealing, and D-Wave’s hybrid solver across increasing graph sizes. For small graphs, all solvers yield the same optimal values. However, quantum annealing performance deteriorates sharply starting from the $(10,70)$ instance, where QA incurs significantly higher cut costs. This degradation is likely due to \emph{embedding overheads, chain break errors, or insufficient annealing schedule expressivity} in larger QUBO encodings. 
Figure~\ref{fig:fig_qa_runtime} shows runtime scaling on D-Wave hardware. QA remains efficient (under 1 second) for small graphs but exhibits a two-order-of-magnitude spike on the $(10,70)$ instance, likely reflecting increased overhead for embedding and retries. In contrast, the hybrid solver maintains stable runtime (10–20 seconds) across all sizes, while consistently achieving better solution quality. These observations highlight a key trade-off between computational time and output accuracy, with hybrid workflows offering robustness as instance size increases.

\begin{figure}[hbtp]
\centering
\begin{minipage}[t]{0.48\textwidth}
    \centering
    \includegraphics[width=\linewidth]{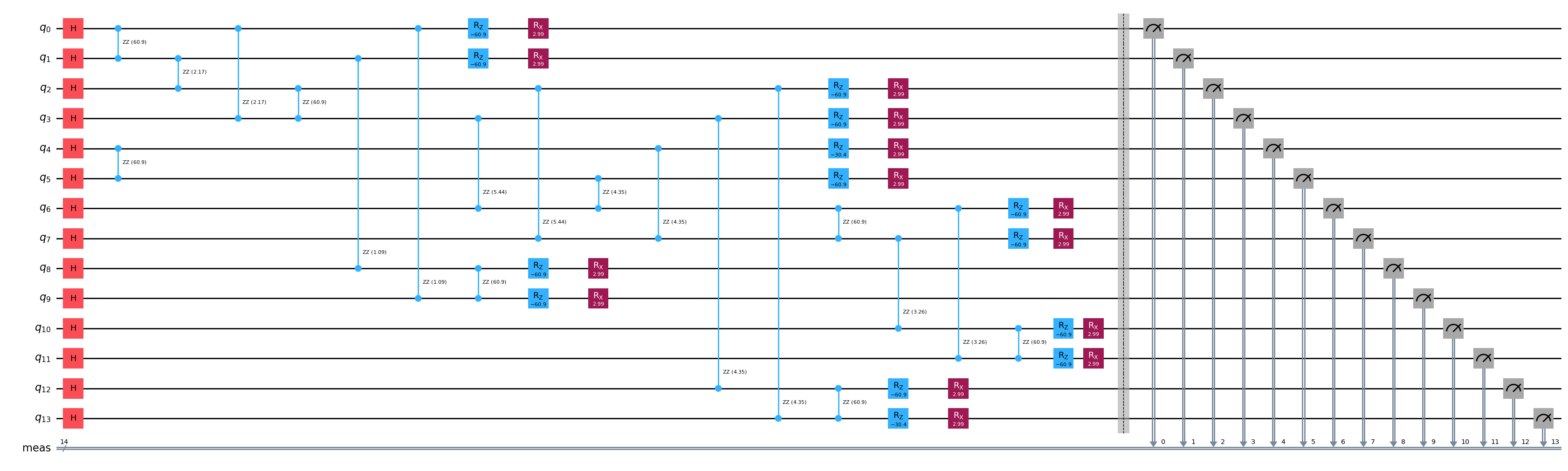}
    \caption{QAOA Circuit Structure ($p = 1$)}
    \label{fig:fig_qaoa_circuit}
\end{minipage}
\hfill
\begin{minipage}[t]{0.48\textwidth}
    \centering
    \includegraphics[width=\linewidth]{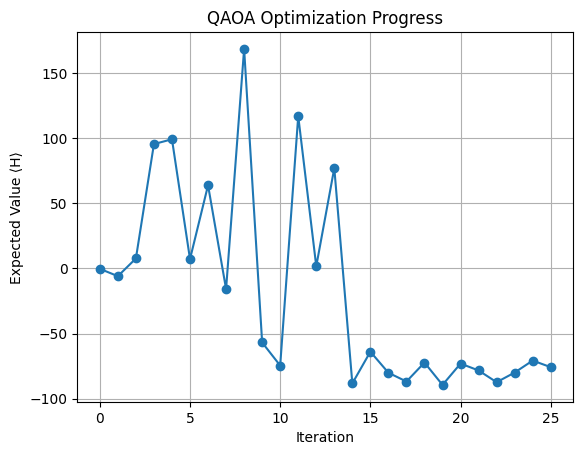}
    \caption{QAOA Training Convergence}
    \label{fig:fig_qaoa_training}
\end{minipage}
\end{figure}

Figure~\ref{fig:fig_qaoa_circuit} shows the QAOA circuit structure for a depth-$p=1$ instance on a small MEMC graph. The ansatz alternates between entangling ZZ-cost layers and parametrized single-qubit mixers. While the structure is amenable to gate-based simulation, circuit width and entanglement depth increase quadratically with instance size, limiting simulation feasibility to $n \lesssim 14$ qubits.

Figure~\ref{fig:fig_qaoa_training} reports the optimization trajectory of the expected cost across 25 classical steps Despite operating in a noise-free environment, the energy landscape remains rugged, leading to unstable convergence and local minima. This reflects known challenges such as barren plateaus in variational algorithms. While convergence is eventually reached, QAOA failed to scale to larger MEMC instances ($k \geq 3$ and $|V| \geq 15$ ) in our setting. Nevertheless, it remains more expressive than photonic workflows in terms of circuit design and model capacity.

\begin{figure}[hbtp]
\centering
\begin{minipage}[t]{0.48\textwidth}
    \centering
    \includegraphics[width=1.1\linewidth]{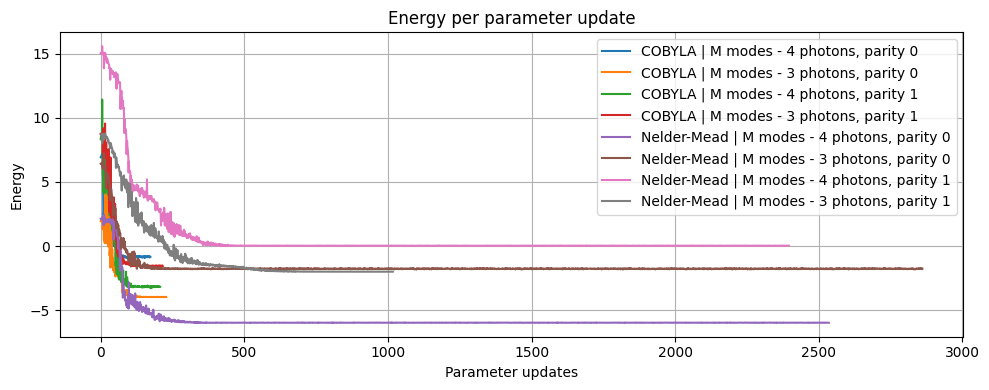}
    \caption{Photonic Energy Convergence}
    \label{fig:fig_photonic_energy}
\end{minipage}
\hfill
\begin{minipage}[t]{0.48\textwidth}
    \centering
    \includegraphics[width=1.2\linewidth]{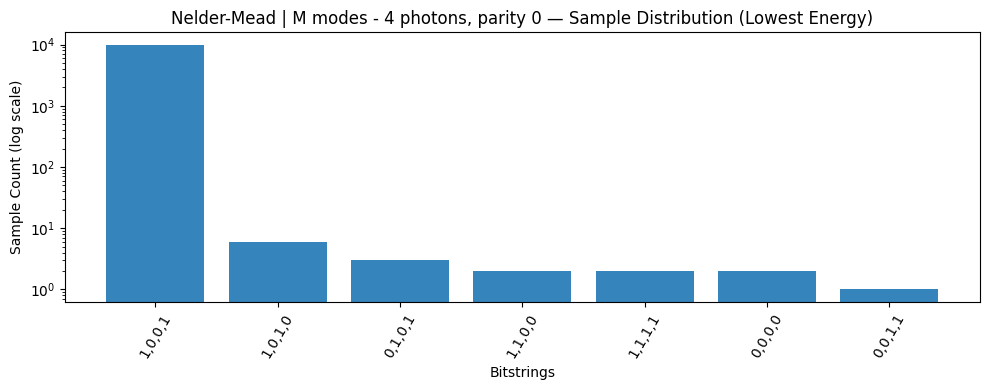}
    \caption{Photonic Bitstring Frequencies}
    \label{fig:fig_photonic_bitstring}
\end{minipage}
\end{figure}

Photonic simulations were conducted using Perceval’s \texttt{GenericInterferometer} and the SLOS backend, combined with classical optimizers. For each parameter configuration, $N = 10000$ measurement shots estimated the QUBO cost. In Figure~\ref{fig:fig_photonic_bitstring}, a toy instance ($|V|=4$, $k=2$) demonstrates strong sampling concentration on the optimal bitstring. Figure~\ref{fig:fig_photonic_energy} shows convergence of QUBO energy using multiple optimizers, with Nelder–Mead yielding greater stability.

Nonetheless, scalability of the photonic workflow remains limited: performance degrades significantly for problems with more than 10 variables or $k \geq 3$. Embedding effects, photon loss, and decoherence are not modeled here. Future directions include testing with qudit encodings or hybrid analog encodings to address mode limitations.
\section{Conclusion}
We benchmarked the MEMC problem across three quantum paradigms under the same QUBO encoding. For small instances ($|V| \leq 10$, $k \leq 2$), all methods—quantum annealing, QAOA, and photonic variational circuits—recovered optimal solutions, confirming baseline compatibility.
Among these, D-Wave showed the highest scalability, solving larger graphs with reasonable solution quality despite embedding overhead. The QUBO formulation, designed with annealing in mind, favored sparse, pairwise interactions, reinforcing D-Wave’s suitability.
QAOA was limited by circuit depth and barren plateaus. Despite convergence in simulation, performance did not extend to medium-scale problems. More advanced designs (e.g., constraint-preserving mixers~\cite{hadfield2019quantum}, parameter concentration~\cite{brady2021bang}) remain to be tested.
Photonic simulations reliably captured optimal bitstrings on toy instances but degrade rapidly beyond $\sim$10 variables. Current parity encodings and the exponential state space are major constraints. Hardware-oriented enhancements (e.g., qudit encodings) are promising.

Overall, Quantum annealing remains the most mature option at scale. Future work will target more expressive encodings, noise-resilient QAOA circuits, and physical deployment of photonic models.
\section{Supplementary Material}
\vspace{-0.75cm}
\begin{table}[htp!]
\centering
\small
\caption{System properties and algorithmic parameters across the three backends.}
\label{tab:exp_params}

\begin{tabular}{p{2.8cm} p{7.5cm}}
\toprule
\multicolumn{2}{c}{\textbf{D-Wave Advantage (eu-central-1)}} \\
\midrule
QPU & Advantage\_system5.4, Pegasus topology (5614 working qubits) \\
Temperature & 16.4 mK \\
Supported Problems & QUBO, Ising \\
Annealing Time & [0.5, 2000] $\mu$s (default 20 $\mu$s) \\
Embedding & Automatic minor-embedding, avg.\ chain length \\
Coupling Ranges & $J \in [-2,1]$, $h \in [-18,15]$ \\
Reads / Postproc & Multiple reads, majority-vote unembedding \\
\midrule
\multicolumn{2}{c}{\textbf{Quandela Perceval (Photonic VQC, simulated)}} \\
\midrule
Backend & SLOS shot-based simulator \\
Ansatz & Generic interferometer (beam splitters + phase shifters) \\
Encoding & Parity-mapped Fock states ($j=0,1$), full/reduced occupancy \\
Parameters & Phases $\{\theta_i,\phi_{tr,i},\phi_i\}$ uniform in $[0,2\pi]$ \\
Optimizers & Powell, COBYLA, Nelder–Mead (max.\ 1000 it.) \\
Loss Function & Variational energy $\langle \psi(\vec{\theta})|H|\psi(\vec{\theta})\rangle$ \\
Samples & 10\,000 per evaluation \\
Postproc & Parity conversion, histogramming of distributions \\
\midrule
\multicolumn{2}{c}{\textbf{IBM Qiskit QAOA (gate-based, simulated)}} \\
\midrule
Backend & Qiskit AerSimulator (statevector + 4000 shots) \\
Ansatz & QAOAAnsatz, depth $p=1$ ($U_C$, $U_M$) \\
Initialization & $\beta_0=\pi/2$, $\gamma_0=\pi/4$ \\
Optimizer & COBYLA, bounds $[0,\pi]$ \\
Objective & $\langle \psi(\vec{\beta},\vec{\gamma})|H|\psi(\vec{\beta},\vec{\gamma})\rangle$ \\
Compilation & Transpilation, preset pass manager (opt.\ level 0) \\
Sampling & All qubits measured, normalized distributions \\
\bottomrule
\end{tabular}
\end{table}
\vspace{-0.75cm}
\paragraph{Reproducibility and scope.}
All implementations use public SDKs (Ocean, Qiskit, Perceval), with parameter ranges listed in Table~\ref{tab:exp_params}. Classical baselines are not included here; they remain essential for competitiveness analysis but outside the scope of this feasibility study.
\paragraph{Contributions and scope.}
This work provides one of the first cross-paradigm studies of the \emph{Minimum Edge Multiway Cut} (MEMC) using a single QUBO formulation across three modalities: D-Wave annealing, gate-based QAOA, and photonic variational circuits. The main contributions are:
(i) a unified encoding enabling consistent benchmarking of embedding overheads and workflow constraints;
(ii) a qualitative analysis of scalability bottlenecks 

(iii) an application-driven link between MEMC and telecom-network resilience;

Unlike prior work focused on native device problems (e.g., Max-Cut on hardware graphs), we benchmark a combinatorial problem with direct industrial motivation. The target is feasibility and methodology, not quantum advantage.

\paragraph{Telecom resilience}
Finding minimum sets of “critical” edges (such that a simultaneous breakdown of these edges would imply a shutdown of a service) is essential for designing “resilient” telecom networks with low probability of failure. This is a hard task since there is a trivial reduction from the NP-hard problem “min edge multiway cut” to this problem.

\end{document}